# On the energy of physical states in QED in the covariant gauge

by

Dan Solomon

Rauland-Borg Corporation

3450 W. Oakton

Skokie, IL USA

Email: dan.solomon@rauland.com

## Abstract

In quantum field theory it is generally assumed that there is a lower bound to the energy of a quantum state. Here, it will be shown that there is no lower bound to the energy of physical states in QED in a manifestly covariant gauge.

1. **<u>Introduction.</u>**

The energy of a normalized state vector $|\Omega\rangle$ is given by $\langle\Omega|H|\Omega\rangle$ where $H$ is the Hamiltonian operator. It is generally assumed that there is a lower bound to the energy of a quantum state which is generally considered to be the energy of the vacuum state. However recent work has shown that this assumption is not necessarily valid. For example, it has been shown that in Dirac's hole theory there exist quantum states with less energy than that of the vacuum state [1,2]. It has also been shown that in Dirac field theory the assumption that the vacuum is the state of minimum energy is in conflict with the requirement of gauge invariance and that the theory will be mathematically inconsistent if the vacuum is required to be the state of minimum energy [2,3,4,5]. In [4] it was shown how one could define a vacuum state which was not the state of minimum energy and that this would result in a mathematically consistent theory. In addition, it has also been shown that the vacuum state is not a minimum energy state for a Dirac-Maxwell field in the temporal gauge [6].

In light of this research the question of whether or not the vacuum state is the state of minimum energy should not be assumed but should be carefully examined and proved or disproved for the quantum theory in question. In this paper we will examine whether or not there is a lower bound to the energy of physical states in QED in a manifestly covariant gauge (or "covariant gauge" for short). It will be shown that there is no lower bound for this situation.

Before proceeding it worth examining further the generally held belief that the vacuum state is the state of minimum energy. How would we prove such a hypothesis for a given quantum theory? The most direct way would be to solve the equation,

$$H|\varphi_n\rangle = E_n|\varphi_n\rangle \tag{1.1}$$

and obtain the eigenstates $|\varphi_n\rangle$ and eigenvalues $E_n$. We could then identify which of the eigenstates $|\varphi_n\rangle$ was the vacuum state and determine if its eigenvalue was less than that of all the other eigenvalues.

For the QED in the covariant gauge the Hamiltonian is given by Eq. (2.4) in the next section. The problem is that, given the complexity of this Hamiltonian, equation (1.1) can't be solved. Except for the simple case of non-interacting field theories this is normally the case. Therefore the widely held belief that the vacuum state is the minimum energy state is merely a *conjecture* with no mathematical proof. As has already been discussed this conjecture has been shown to be wrong for Dirac's hole theory[1,2] and for a Dirac-Maxwell field in the temporal gauge[6]. With this in mind we will proceed to show that there is no lower bound to the energy of physical states for QED in the covariant gauge.

2. **QED in the covariant gauge.**

A covariant gauge is a gauge in which a Lorentz transformation does not change the gauge condition. Physical states are those states that satisfy certain constraints so that the gauge fields satisfy the standard Maxwell equations. The following discussion of the covariant gauge is based on the work of Haller and Lim-Lombridas[7] and the notation of [7] will be used.

The Lagrangian for QED in the covariant gauge is given by [7],

$$L = -\frac{1}{4} F_{\mu\nu} F^{\mu\nu} - j_\mu A^\mu - G(\partial_\mu A^\mu) + \frac{1}{2}(1-\gamma)G^2 + \bar{\psi}(i\gamma_\mu \partial^\mu - m)\psi \quad (2.1)$$

where $j_\mu = e\bar{\psi}\gamma_\mu \psi$, $F_{\mu\nu} = \partial_\mu A_\nu - \partial_\nu A_\mu$, $\gamma$ is a constant, and $G$ is the gauge fixing field. From this Lagrangian we can derive the Euler-Lagrange equations of motion for the gauge fields. Some of these are [7],

$$\partial_0 \mathbf{E} - \nabla \times \mathbf{B} + \mathbf{j} = -\nabla G \quad (2.2)$$

and

$$\nabla \cdot \mathbf{E} - j_0 = -\partial_0 G \quad (2.3)$$

The Hamiltonian is given by,

$$H = H_0 + H_I \quad (2.4)$$

where,

$$H_0 = \int \mathcal{H}_0(\mathbf{x}) d^3x; \quad H_I = \int \mathcal{H}_I(\mathbf{x}) d^3x \quad (2.5)$$

with,

$$H_0(\mathbf{x}) = \frac{1}{2}\mathbf{\Pi}\cdot\mathbf{\Pi} + \frac{1}{4}F_{ij}F^{ij} + G\nabla\cdot\mathbf{A} + A^0\nabla\cdot\mathbf{\Pi} - \frac{1}{2}(1-\gamma)G^2 + \psi^\dagger(-i\boldsymbol{\alpha}\cdot\nabla+\beta m)\psi \quad (2.6)$$

and

$$H_I(\mathbf{x}) = j_0 A^0 - \mathbf{j}\cdot\mathbf{A} \quad (2.7)$$

where $\mathbf{\Pi} = -\mathbf{E}$ where $\mathbf{E}$ is the electric field. In addition to the above, a number of useful relationships involving the field operators, including equal time commutation relationships, are provided in [7].

The operators act on a Fock space $\{|h\rangle\}$. Now the operator equations (2.2) and (2.3) do not obey Maxwell's equations because of the presence of the term $G$. Therefore in order that Maxwell's equations are obeyed we designate a set of states $\{|v\rangle\}$ which are called the "physical subspace" or physical states. The set $\{|v\rangle\}$ is a subset of $\{|h\rangle\}$ and satisfies the condition,

$$\Omega(\mathbf{k})|v\rangle = 0 \quad (2.8)$$

where $\Omega(\mathbf{k})$ is defined in [7]. When this condition is satisfied then, as is shown in [7], the following relationships hold,

$$\langle v_2|\partial_0 G|v_1\rangle = \langle v_2|G|v_1\rangle = 0 \quad (2.9)$$

Therefore, when the physical states are used, Maxwell's equations will be obeyed.

### 3. Energy of physical states.

In this section we will consider the question as to whether or not there is a lower bound to the energy of the physical states. That is, does there exist a normalized physical state $|v_{vac}\rangle$ such that,

$$\langle v|H|v\rangle \geq \langle v_{vac}|H|v_{vac}\rangle \text{ for all normalized physical states } |v\rangle \quad (3.1)$$

First assume that there exists a normalized physical state $|v'\rangle$ for which the quantity $\langle v'|\nabla\cdot\mathbf{j}(\mathbf{x})|v'\rangle$ is non-zero for some region of space, i.e.,

$$\langle v'|\nabla\cdot\mathbf{j}(\mathbf{x})|v'\rangle \neq 0 \quad (3.2)$$

How do we know that such a state exists? The quantity $\langle v'|\nabla\cdot\mathbf{j}(\mathbf{x})|v'\rangle$ is the expectation value of the divergence of the current. And we know that there are many situations in the

real world where the divergence of the current is non-zero over some region of space. Therefore, if our theory is a correct model of the real world there must be some states for which (3.2) holds over some region of space.

Next define the unitary operator,

$$U = e^{-iC}; \quad U^\dagger = e^{+iC} \tag{3.3}$$

where,

$$C = \int \mathbf{E} \cdot \nabla \chi d^3 x \tag{3.4}$$

and where $\chi(\mathbf{x})$ is an arbitrary real valued function with reasonable boundary conditions at infinity so that we can integrate by parts. Note that we have used the fact that $C^\dagger = C$ (since $\mathbf{E}$ is real) to obtain (3.3). Next define the state,

$$|\xi\rangle = U|v'\rangle \tag{3.5}$$

where $|v'\rangle$ is a normalized physical state. Note that $|\xi\rangle$ is normalized due to the fact that $U^\dagger U = 1$. It is shown in the Appendix that $[U, \Omega(\mathbf{k})] = 0$. Using this, along with (2.8), we obtain,

$$\Omega(\mathbf{k})|\xi\rangle = \Omega(\mathbf{k})U|v'\rangle = U\Omega(\mathbf{k})|v'\rangle = 0 \tag{3.6}$$

therefore $|\xi\rangle$ is a physical state. Next we want to determine the energy $\langle\xi|H|\xi\rangle = \langle v'|U^\dagger H U|v'\rangle$. In order to do this we will use the Baker-Campell-Hausdorff relationships which state that,

$$e^{+O_A} O_B e^{-O_A} = O_B + [O_A, O_B] + \frac{1}{2}[O_A, [O_A, O_B]] + \ldots \tag{3.7}$$

Use this along with (3.5) and (3.3) to obtain,

$$\langle\xi|H|\xi\rangle = \langle v'|\left(H + [iC, H] + \frac{1}{2}[iC, [iC, H]] + \ldots\right)|v'\rangle \tag{3.8}$$

To evaluate this we must first determine $[H, \mathbf{E}]$. This can be obtained by using the various commutation relationships given in [7]. However it is much simpler to refer to (2.2) and make the usual substitution $\partial_0 \to i[H, \ ]$ to obtain,

$$i[H, \mathbf{E}] = \nabla \times \mathbf{B} - \mathbf{j} - \nabla G \tag{3.9}$$

(Note, if we had used the relationships of [7] we would obtain the some result). From this we obtain,

$$[iC, H] = -\int (\nabla \times \mathbf{B} - \mathbf{j} - \nabla G) \cdot \nabla \chi d^3 x = -\int \chi (\nabla \cdot \mathbf{j} + \nabla^2 G) d^3 x \qquad (3.10)$$

where we have integrated by parts and used the identity $\nabla \cdot (\nabla \times \mathbf{B}) = 0$ to obtain the final result. It is shown in the Appendix that $[G, \mathbf{E}]$ and $[\mathbf{j}, \mathbf{E}]$ are both zero. Therefore,

$$[iC, [iC, H]] = 0 \qquad (3.11)$$

Use these results in (3.8) to obtain,

$$\langle \xi | H | \xi \rangle = \langle v' | \left( H - \int \chi (\nabla \cdot \mathbf{j} + \nabla^2 G) d^3 x \right) | v' \rangle \qquad (3.12)$$

This can be rewritten as,

$$\langle \xi | H | \xi \rangle = \langle v' | H | v' \rangle - \int \chi \left( \langle v' | \nabla \cdot \mathbf{j} | v' \rangle d^3 x \right) d^3 x - \left( \int \chi \nabla^2 \langle v' | G | v' \rangle d^3 x \right) \qquad (3.13)$$

Next use the fact that $|v'\rangle$ is a physical state so that $\langle v' | G | v' \rangle = 0$ to obtain,

$$\langle \xi | H | \xi \rangle = \langle v' | H | v' \rangle - \int \chi \left( \langle v' | \nabla \cdot \mathbf{j} | v' \rangle d^3 x \right) \qquad (3.14)$$

Note in the above expression the function $\chi(\mathbf{x})$ does not appear in the quantity $\langle v' | H | v' \rangle$. Therefore, due to the fact that $\langle v' | \nabla \cdot \mathbf{j}(\mathbf{x}) | v' \rangle$ is nonzero we can always find a $\chi(\mathbf{x})$ so that $\langle \xi | H | \xi \rangle$ is an arbitrarily large negative number. For example, let,

$$\chi(\mathbf{x}) = f \langle v' | \nabla \cdot \mathbf{j}(\mathbf{x}) | v' \rangle \qquad (3.15)$$

where $f$ is positive number. Use this in (3.14) to obtain,

$$\langle \xi | H | \xi \rangle = \langle v' | H | v' \rangle - f \int \langle v' | \nabla \cdot \mathbf{j} | v' \rangle^2 d^3 x \qquad (3.16)$$

The integral must be positive. Therefore as $f \to \infty$ then $\langle \xi | H | \xi \rangle \to -\infty$. Therefore there is no lower bound to the energy of a physical state in the covariant gauge. This is consistent with the results of [6] where it was shown that there was no lower bound to the energy of physical states in the temporal gauge.

**Appendix**

Refer to [7] to obtain,

$$\Omega(\mathbf{k}) = a_Q(\mathbf{k}) + \frac{j_0(\mathbf{k})}{2k^{3/2}} \qquad (A.1)$$

$$G(\mathbf{x}) = i\sum_{\mathbf{k}} \sqrt{k} \left[ a_Q(\mathbf{k}) e^{i\mathbf{k}\cdot\mathbf{x}} - a_Q^*(\mathbf{k}) e^{-i\mathbf{k}\cdot\mathbf{x}} \right] \tag{A.2}$$

and,

$$E_i(\mathbf{x}) = i\sum_{\mathbf{k}} \varepsilon_i^n(\mathbf{k}) \sqrt{\frac{k}{2}} \left[ a_n(\mathbf{k}) e^{i\mathbf{k}\cdot\mathbf{x}} - a_n^\dagger(\mathbf{k}) e^{-i\mathbf{k}\cdot\mathbf{x}} \right] + i\sum_{\mathbf{k}} \frac{k_i}{\sqrt{k}} \left[ a_Q(\mathbf{k}) e^{i\mathbf{k}\cdot\mathbf{x}} - a_Q^*(\mathbf{k}) e^{-i\mathbf{k}\cdot\mathbf{x}} \right] \tag{A.3}$$

From [7] $a_Q(\mathbf{k})$ and $a_Q^*(\mathbf{q})$ obey the following unusual commutation relationship,

$$\left[ a_Q(\mathbf{k}), a_Q^*(\mathbf{q}) \right] = \left[ a_Q(\mathbf{k}), a_Q(\mathbf{q}) \right] = \left[ a_Q^*(\mathbf{k}), a_Q^*(\mathbf{q}) \right] = 0 \tag{A.4}$$

In addition,

$$\left[ a_n(\mathbf{k}) \text{ or } a_n^\dagger(\mathbf{k}), a_Q(\mathbf{q}) \text{ or } a_Q^*(\mathbf{q}) \right] = 0 \tag{A.5}$$

Also the gauge field operators commute with the fermion field operators so that, $a_Q(\mathbf{k})$ and $a_Q^*(\mathbf{k})$ commute with the fermion field operators so that,

$$\left[ a_Q^*(\mathbf{k}) \text{ or } a_Q(\mathbf{k}) \text{ or } a_n(\mathbf{k}) \text{ or } a_n^\dagger(\mathbf{k}), j_\mu \right] = 0 \tag{A.6}$$

Therefore $\left[ \Omega(\mathbf{k}), E_i \right] = 0$  This yields $\left[ \Omega(\mathbf{k}), C \right] = 0$ which yields $\left[ \Omega(\mathbf{k}), U \right] = 0$. In addition we have $\left[ G(\mathbf{x}), \mathbf{E}(\mathbf{y}) \right] = 0$ and $\left[ \mathbf{j}(\mathbf{x}), \mathbf{E}(\mathbf{y}) \right] = 0$.